\documentclass[aps,prb,twocolumn,superscriptaddress, longbibliography]{revtex4-1}
\usepackage{amsmath,amssymb,amsfonts,upgreek}
\usepackage{color}
\usepackage{graphicx}
\usepackage{setspace}
\usepackage{bm}
\usepackage[export]{adjustbox}
\usepackage{hyperref}
\usepackage{dsfont}
\usepackage{braket}
\usepackage{changes}

\newcommand{\rone}{$R_1^+$}

\begin{document}

\title{Interplay between breathing-mode distortions and magnetic order in rare-earth nickelates from \textit{ab initio} magnetic models}

\author{Danis I. Badrtdinov}
\affiliation{Theoretical Physics and Applied Mathematics Department, Ural Federal University, 620002 Yekaterinburg, Russia}
\affiliation{Center for Computational Quantum Physics, Flatiron Institute, 162 Fifth avenue, New York, NY 10010, USA}
\author{Alexander Hampel}
\email{ahampel@flatironinstitute.org}
\affiliation{Center for Computational Quantum Physics, Flatiron Institute, 162 Fifth avenue, New York, NY 10010, USA}
\author{Cyrus E. Dreyer}
\affiliation{Center for Computational Quantum Physics, Flatiron Institute, 162 Fifth avenue, New York, NY 10010, USA}
\affiliation{Department of Physics and Astronomy,   Stony Brook University,  Stony Brook, New York, 11794-3800, USA}

\date{\today}

\begin{abstract}

We use density-functional theory calculations to explore the magnetic properties of perovskite rare-earth nickelates, $\mathcal{R}$NiO$_3$, by constructing microscopic magnetic models containing all relevant exchange interactions via Wannierization and Green's function techniques. These models elucidate the mechanism behind the formation of antiferromagnetic order with the experimentally observed propagation vector, and explain the reason  previous DFT plus Hubbard $U$ calculations favored ferromagnetic order. We perform calculations of magnetic moments and exchange-coupling parameters for different amplitudes of the \rone{} breathing mode distortion, which results in expanded and compressed NiO$_6$ octahedra. We find that the magnetic moment vanishes for the ``short bond'' nickels, i.e., the ones in the compressed octahedra. The inclusion of spin-orbit coupling demonstrates that the magnetic anisotropy is very small, while the magnetic moment of the short bond nickel atoms tend to  zero even for the noncollinear case. Our results provide a clear picture of the trends of the magnetic order across the nickelate series and give insights into the coupling between magnetic order and structural distortions.

\end{abstract}

\maketitle

\section{Introduction}

Rare-earth (RE) nickelate perovskites, with the chemical formula $\mathcal{R}$NiO$_3$, have received significant attention  due to their rich phase diagram. As a function of the ionic size of RE element $\mathcal{R}$~\cite{hampel2017, varignon2017, catalano2018, lu2017},  a combined structural and metal-insulator transition (MIT) occurs with temperature, as well as  the formation of long-range magnetic order. Both can be tuned by  external pressure~\cite{obradors1993}, strain~\cite{catalano2014} and reduced dimensionality~\cite{boris2011}. This tunability  paves the way for engineering  nanoelectronic and spintronic devices~\cite{cao2015, shi2014, cheong2007}.

\begin{figure}[t]
    \centering
     \includegraphics[width=1\linewidth]{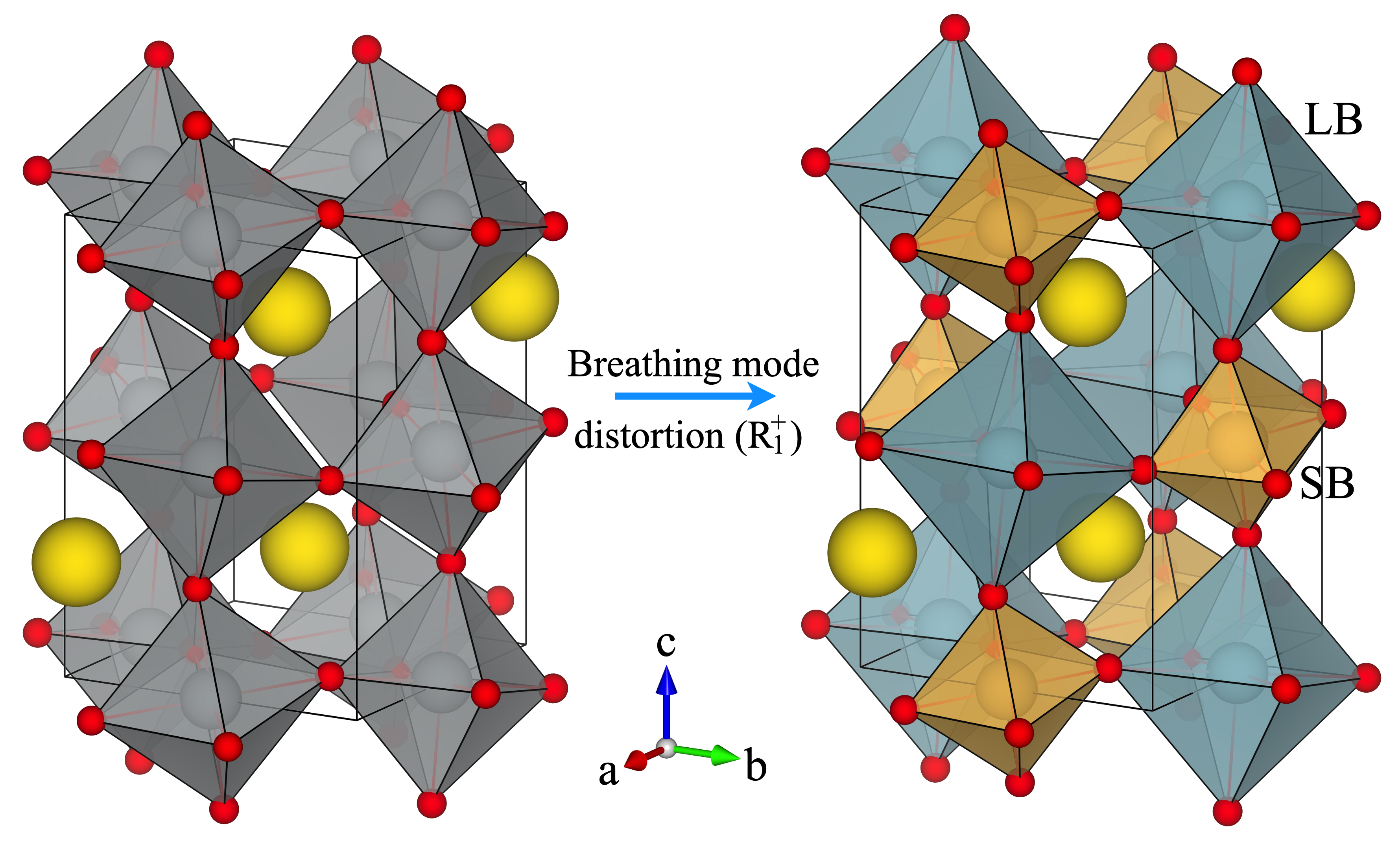}
    \caption{ Evolution of crystal structure of $\mathcal{R}$NiO$_3$ under the breathing mode distortion \rone{}, which leads to formation of short bond (SB) and long  bond (LB) NiO$_6$ octahedra. The rare-earth element $\mathcal{R}$ is denoted by yellow spheres.  Crystal structures are visualized using the VESTA software~\cite{vesta}.
    }
    \label{fig:structure}
\end{figure}

The structural  transition, which occurs in all RE nickelates except LaNiO$_3$,  is characterized by a symmetry lowering from the orthorhombic $Pbnm$  to the monoclinic $P2_1/n$ spacegroup upon cooling; it can be described mainly by a ``breathing-mode'' distortion, which creates expanded and compressed NiO$_6$ octahedra arranged in a three dimensional checkerboard pattern. We refer to the two different Ni sites that result as residing in either the long bond (LB) or short bond (SB) octahedra~\cite{alonso1999} (Fig.~\ref{fig:structure}).  

This distortion strongly modifies the electronic configuration and corresponding magnetic moments of nickel atoms. One of the proposed descriptions of the resulting electronic structure~\cite{Mazin:2007} suggests that the initial electronic configuration of the Ni atoms, i.e., $3d^7$ with spin $S = \frac{1}{2}$, transforms  to two charge-disproportionated  Ni sublattices  $3d^6 \, (S =0)$ and $3d^8 \, (S = 1)$ under the distortion. However, this atomic picture does not consider the strong hybridization between Ni-$3d$ and O-$2p$ orbitals. Hence, later studies introduced a more complete description of the electronic configuration including oxygen (ligand) ``holes,'' denoted as  $\underline{L}$~\cite{johnston2014, green2016}, and describing the connected structural-MIT in nickelates as a site-selective Mott transition~\cite{park2012, lau2013, johnston2014, subedi2015, haule2017, peil2019}. The  resulting SB and LB octahedra are described by $d^8\underline{L}^{0}$ and $d^8\underline{L}^{2}$ states with total spins tending toward $S = 0$ and $S=1$, respectively. Conventional density functional theory (DFT) based methods also support this interpretation, showing different magnetic moments of Ni$_{\rm SB}$ and Ni$_{\rm LB}$ atoms, while demonstrating nearly  equal integrated charges around both Ni atoms inside the projector augmented wave (PAW) spheres~\cite{park2012, hampel2017}. However, such an analysis will depend on the size of the augmentation region~\cite{kresse_PAW}, and cannot fully capture the hybridization with the oxygens; a more complete picture may be obtained, e.g., via constructing Wannier functions for the (possibly hybridized) Ni $3d$ states \cite{varignon2017}.

At low temperatures, RE nickelates (again with the exception of LaNiO$_3$) exhibit an antiferromagnetic (AFM) insulating ground state. The magnetic structure has a propagation vector of $\mathbf{q} = (\frac{1}{4}, \frac{1}{4},\frac{1}{4})$ (in pseudocubic notation),  which was probed by neutron scattering for bulk systems~\cite{garcia1994, alonso1999, gawryluk2019, Munoz:2009,FernandezDiaz:2001} and resonant soft X-ray scattering (RIXS) for thin film heterostructures~\cite{scagnoli2006, lu2018}. The exact arrangement of magnetic moments in the systems is still under debate and strongly depends on the interpretation of experimental data~\cite{haule2017, catalano2018}. Particularly, neutron data suggests a collinear magnetic ordering, where the magnetic moment of SB nickel atoms vanish~\cite{Munoz:2009,FernandezDiaz:2001,haule2017}. One candidate for the ground-state magnetic configuration is a T-type AFM (T-AFM) structure \cite{catalano2018,hampel2017}, depicted in Fig.~\ref{fig:Magnetic_ordering}(a). In this structure, the LB moments are arranged in an ``up-up-down-down'' ($\uparrow \uparrow \downarrow \downarrow$) pattern in all three Cartesian directions, and the SB moments tend to zero with increasing \rone{} amplitude. 

On the other hand, the corresponding RIXS spectra have been interpreted as a non-collinear (NCL) arrangement of spins with clearly non-zero SB magnetic moments~\cite{scagnoli2006, lu2018}, also involving the $\uparrow \uparrow \downarrow \downarrow$  pattern in all three Cartesian directions [see Fig.~\ref{fig:Magnetic_ordering}(b)]. Here, all moments lie in the $ac$ plane, with their direction reversed upon moving along [1, 0, 1]. This magnetic configuration can be considered as an orthogonal spin spiral. More importantly, the nearest neighbor SB and LB  magnetic moments are perpendicular to each other, which can weaken interactions between them. 

Our study is aimed at understanding the formation of these proposed complex magnetic orderings on a microscopic level in the RE nickelates. To this end we performed a systematic DFT plus Hubbard $U$ (DFT+$U$) analysis of the magnetic properties of a series of RE nickelates: PrNiO$_3$, SmNiO$_3$ and LuNiO$_3$. We will consider the collinear T-AFM and NCL order [Fig.~\ref{fig:Magnetic_ordering}(a) and (b)], as well as the ferromagnetic (FM) case, which was shown to have the lowest energy in previous DFT+$U$ studies~\cite{hampel2017,varignon2017}. Spin models are constructed by obtaining Wannier functions for the hybridized Ni($3d$)-O($2p$) orbitals, and then using the local force theorem approach \cite{liechtenstein1987, mazurenko2005} to extract exchange couplings. We analyze the evolution of the magnetic interactions under the \rone{} structural breathing mode distortion. Our calculations reveal strong ferromagnetic nearest-neighbor and weaker antiferromagnetic next-nearest-neighbor couplings, which are induced due to magnetically active Ni $d_{z^2}$ and $d_{x^2-y^2}$ states. Competition between these couplings plays the dominant role in determining the formation of magnetic order in $\mathcal{R}$NiO$_3$ systems and explains why previous theoretical calculations favored ferromagnetic ordering. The inclusion of spin-orbit coupling demonstrates that the magnetic anisotropy is very small, while the magnetic moments of SB nickel atoms tend to zero for finite \rone{} amplitude even in the NCL case.

The remainder of this paper is organized as follows. Sec.~\ref{sec:theory} covers the methods and theories used in our calculations. In Sec.~\ref{sec:results}, we present our main results for magnetic moments and interactions of selected rare-earth nickelate systems: PrNiO$_3$, SmNiO$_3$ and LuNiO$_3$.  We discuss the implications of our results on the proposed magnetic structures in Sec.~\ref{sec:discussion}.  Finally, Sec.~\ref{sec:conc} concludes the paper.

%

\section{Theoretical framework}
\label{sec:theory}

\subsection{Formalism}

 We construct the following spin model, which describes energetics of individual spins ${\bf S}_{i} $ on nickel atoms interacting via isotropic exchange couplings $J_{ij} $: 
\begin{equation}
\hat {\mathcal{H}}^{\text{spin}} = \sum_{i>j} J_{ij} \hat {{\bf S}}_{i} \hat {{\bf S}}_{j}.
\label{eq:spin_ham}
\end{equation}
The  exchange integrals  can be calculated using the local force theorem approach~\cite{liechtenstein1987, mazurenko2005}
\begin{equation}
\begin{split}
J_{ij} &=  \frac{1}{2 \pi S^2} \sum \limits_{m, m^{\prime},  n, n^{\prime}} \times \\
  & {\rm Im}   \int \limits_{-\infty}^{E_F} d \epsilon \left[ \Delta^{m m^{\prime}}_i G^{m^{\prime} n}_{ij \downarrow} (\epsilon) \Delta^{n n^{\prime}}_j G^{n^{\prime} m}_{ji \uparrow} (\epsilon) \right],
\label{eq:Exchange}
\end{split}
\end{equation}
where $m, m^{\prime},  n, n^{\prime}$ are orbital indices running over Ni 3$d$ $(l = 2)$ states, $E_F$ is the Fermi energy, $S = \frac{1}{2}$ is the spin quantum number~\footnote{For simplicity, we use spin quantum number  $S = \frac{1}{2}$ in Eq.~(\ref{eq:Exchange}) during the whole range of breathing distortions. It was revealed that very large breathing distortion amplitudes are needed to achieve spins $0$ and $1$ for SB and LB magnetic moments~\cite{johnston2014, green2016}, which is much higher than our considered cases},  $\Delta^{m m^{\prime}}_i = H^{m m^{\prime}}_{ii \uparrow} - H^{m m^{\prime}}_{ii \downarrow}$ is the on-site potential, and $G$ is the single-particle Green's functions
\begin{eqnarray}
G^{n n^{\prime}}_{ij \sigma}(\epsilon) = \frac{1}{N_{\textbf{k}}} \sum \limits_{\mathbf{k},l}^M \frac{c^{nl}_{i \sigma}(\mathbf{k}) c^{n^{\prime}l^*}_{j \sigma}(\bf{k})}{\epsilon - E^{l}_{\sigma}(\mathbf{k})}.
\label{eq:Green}
\end{eqnarray} 
Here $N_{\textbf{k}}$ is the number of $k$ points in the first Brillouin zone (BZ), $c^{n^\prime l*}_{i \sigma}(\mathbf{k})$ [$c^{nl}_{i \sigma}(\mathbf{k})$] is the creation [annihilation] operator for the $l^{\rm th}$ eigenstate in the basis of Wannier functions (see Sec.~\ref{sec:comp}), with eigenvalue $E^{l}_{\sigma}(\mathbf{k})$. The summation in Eq.~(\ref{eq:Green}) runs over  all Ni(3$d$) and O(2$p$) orbitals; since $n$ and $n^\prime$ only consist of Ni($3d$) orbitals, this corresponds to projecting the Wannier functions onto Ni($3d$) states, thus capturing the overlap between Ni($3d$) and O($2p$).

The magnetic moment of a selected nickel site can also be evaluated as
\begin{eqnarray}
m_{i }=  -\frac{1}{ \pi } {\rm Im}   \int \limits_{-\infty}^{E_F} d \epsilon&\,  \mathrm{Tr}_{n} \left[  G^{n n}_{ii \uparrow}  (\epsilon) -  G^{n n}_{ii \downarrow} (\epsilon)  \right].  
\label{eq:Moment_from_denisty}
\end{eqnarray} 
Eq.~(\ref{eq:Moment_from_denisty}) includes in the Ni moments the contribution from the overlap with oxygen. If we define a different set of Green's functions by allowing $n$ and $n^{\prime}$ in Eq.~(\ref{eq:Green}) to run over all of the Wannier functions (not just nickel $3d$), then Eq.~(\ref{eq:Moment_from_denisty}) corresponds to the total magnetic moment per site, equivalent to integrating over the density of states near the Fermi level; by limiting the trace over just the Ni(3$d$) or just O(2$p$) Wannier functions, we can obtain separate moments on the nickels and oxygens, which may be compared to those integrated within the PAW spheres~\cite{kresse_PAW}, as discussed in the next section.

\subsection{Computational approach \label{sec:comp}}

We calculated the exchange couplings  in Eq.~(\ref{eq:Exchange}) using DFT with the generalized-gradient approximation  exchange-correlation functional of Perdew, Burke, and Ernzerhof~\cite{pbe96} and the projector augmented wave (PAW) method\cite{Blochl1994} as implemented in the Vienna ab initio Simulation Package (\texttt{VASP})~\cite{vasp1, vasp2}.  For the rare-earth atoms, we used PAW potentials corresponding to a $3+$ valence state with $f$-electrons frozen into the core;  for Ni, the 3$p$ semi-core states were included as valence electrons. We set the energy cutoff of the plane-wave basis to 550\,eV and the energy convergence criteria to 10$^{-8}$\,eV. 

Correlation effects were taken into account on the static mean-field level using the DFT+$U$ method~\citep{anisimov1991}, where we considered an effective on-site Coulomb $U$ and  Hund's exchange $J_H$ interactions in the rotationally invariant form~\cite{liechtenstein1995}. A value of $J_H=1$ eV  can be used for compounds with 3$d$ orbitals~\cite{danis2016, danis2019}, while the on-site Coulomb parameter depends on the specific system. Previous results showed that the best agreement with experiment can be obtained using a relatively small value of $U$ = 2 eV~\cite{hampel2017, varignon2017, mercy2017}. Therefore, we used $U$ = 2 eV and $J_\text{H}$ = 1 eV for the majority of our DFT+$U$ calculations, and comment on how a larger $U$ ($U$ = 5 eV) may effect the results. 

\begin{figure}[!t]
    \centering
     \includegraphics[width=0.8\linewidth]{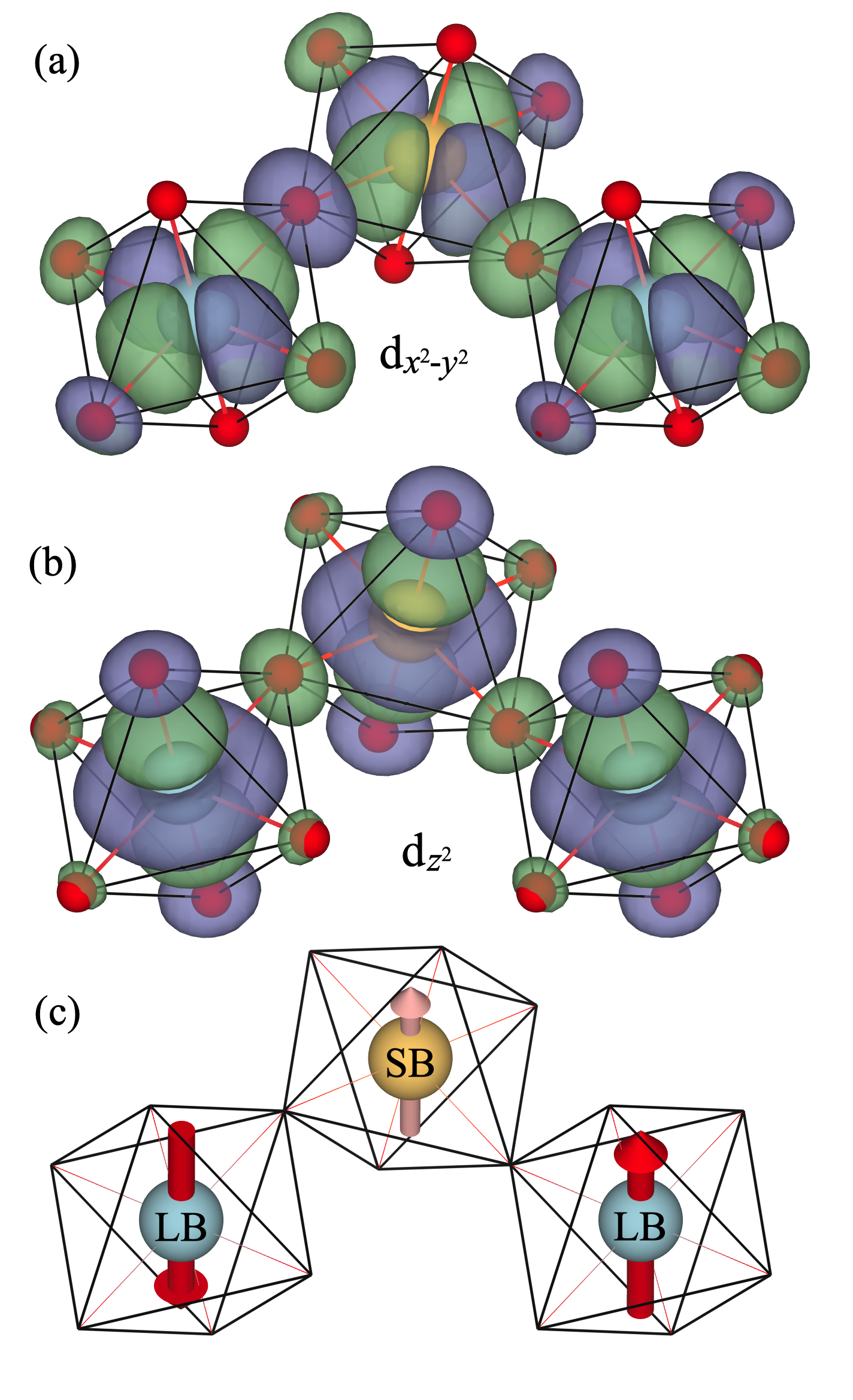}
    \caption{ Maximally localized Wannier functions of the long-bond (LB) and short-bond (SB) nickel atoms. (a) corresponds to $d_{x^2-y^2}$ and (b) to $d_{z^2}$ orbitals for nearest neighbor LB -- SB -- LB nickel atoms, calculated for the T-AFM PrNiO$_3$ system at the experimental breathing mode amplitude \rone{} = 0.045 \AA. Only spin majority orbitals at each nickel atoms are represented for clarity, spin moments are shown in (c).
    }
    \label{fig:WF}
\end{figure}

From the calculated electronic structure, maximally-localized Wannier functions~\cite{marzari1997} were generated using the \texttt{wannier90} package~\cite{pizzi2020}. We constructed Wannier functions for all occupied and unoccupied Ni 3$d$  and O 2$p$ states. This ensurs that the obtained Wannier functions for both spin channels are well localized (Fig.~\ref{fig:WF}), and we can use them as an atomic-like basis for Eqs.~(\ref{eq:Exchange}) and (\ref{eq:Green}). Note, that this choice is different to Ref.~\onlinecite{varignon2017}, were Wannier functions were only constructed for the occupied states. Since the $f$ electrons of the RE elements were frozen into core states, we have neglected their magnetic ordering  which only occurs at very low temperatures~\cite{FernandezDiaz:2001,Munoz:2009}.

To examine different magnetic orders we considered two types of cells:  the primitive monoclinic  cell containing 20 atoms, and a cell enlarged in the $a$ and $c$ directions containing 80 atoms (necessary for the AFM orderings); an 8$\times$8$\times$6 and 4$\times$6$\times$4 Monkhorst-Pack mesh for  Brillouin-zone integration were used for the 20 and 80 atom cells, respectively.  The number of $k$-points within the first Brillouin zone for the evaluation of the Green's function [Eq.~(\ref{eq:Green})] and energy points for integration over a semicircular contour in the upper half of the complex energy plane [Eqs.~(\ref{eq:Exchange}) and (\ref{eq:Moment_from_denisty})] were chosen to reach numerical convergence of exchange couplings within 0.1 meV.

For a systematic study of structural distortion, we used a symmetry-based mode decomposition analysis~\cite{perez_mato2010} where we varied the amplitude of specific modes expressed in terms of  irreducible representations of the parent  (cubic perovskite) $Pm\overline{3}m$ structure. The most significant modes that are present in the RE nickelates include octahedral rotation modes $M_3^{+}$ (in-phase) and  $R_4^{+}$ (out-of-phase), and the breathing mode distortion $R_1^{+}$~\cite{Balachandran:2013}. The breathing mode, which is responsible for the formation of SB and LB octahedral (Fig.~\ref{fig:structure}), is the distortion most strongly coupled to the magnetic structure, so it will be the focus of this study. The main effect of the different RE atoms is to change the amplitude of the octahedral rotation modes. Specifically, smaller radii REs correspond to larger rotations. Thus the range in RE atoms in this study  demonstrates the influence of octahedral rotations on the magnetic ordering.~\cite{hampel2019}.

In our calculations, we start from the high-temperature orthorhombic $Pbnm$ structure, relaxed with non-magnetic DFT (with $U=0$)~\cite{hampel2017}; in this case,  SB and LB octahedra have the same volume. We then gradually increase the $R_1^{+}$ amplitude, calculating how the exchange couplings and magnetic moments are modified by the distortion. Thus we can study the evolution of magnetic characteristics with the transition from $Pbnm$ to the low temperature insulating monoclinic phase $P2_1/n$.

\begin{figure}[!t]
    \centering
     \includegraphics[width=1\linewidth]{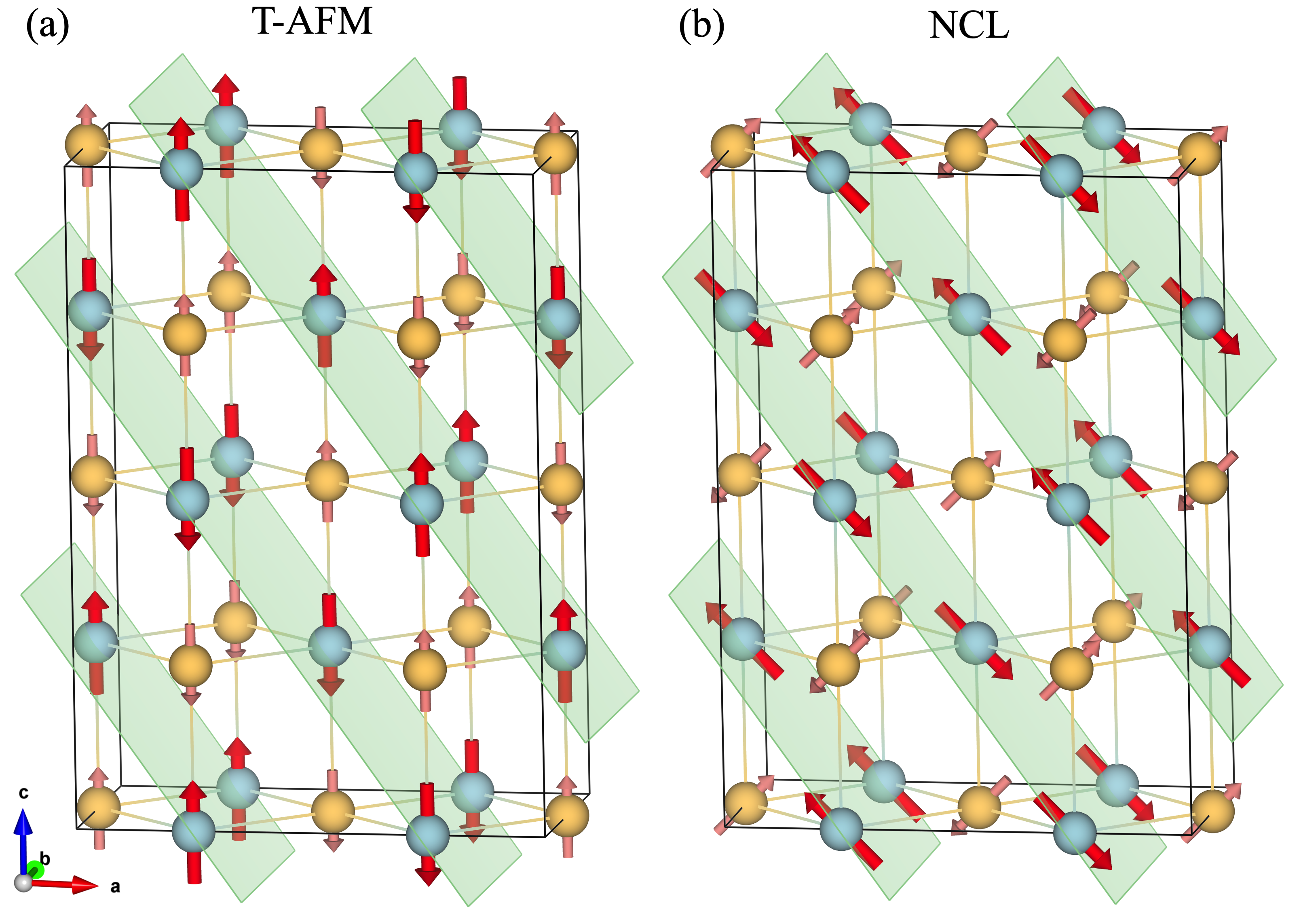}
    \caption{(a) T-type antiferromagnetic (AFM) and (b) noncolinear (NCL) magnetic  configurations as a possible ground state magnetic orderings in $\mathcal{R}$NiO$_3$ with experimental propagation vector $\mathbf{q} = (\frac{1}{4}, \frac{1}{4},\frac{1}{4})$. Yellow and blue spheres correspond to Ni$_{\text{SB}}$ and Ni$_{\text{LB}}$ atoms respectively, which have smaller and larger magnetic moments corresponding to the length of the red arrows. 
    }
    \label{fig:Magnetic_ordering}
\end{figure}

\section{Results}
\label{sec:results}

\subsection{Magnetic moment analysis}

We begin with the magnetic moments of PrNiO$_3$, SmNiO$_3$ and LuNiO$_3$, assuming FM or T-AFM ordering, using the on-site Green's functions [Eq.~(\ref{eq:Moment_from_denisty})] in the Wannier basis. This procedure will correctly capture the hybridization effect between nickel and oxygen states and the corresponding distribution of magnetic moment density~\cite{varignon2017}.

\begin{figure}[!t]
    \centering
     \includegraphics[width=1\linewidth]{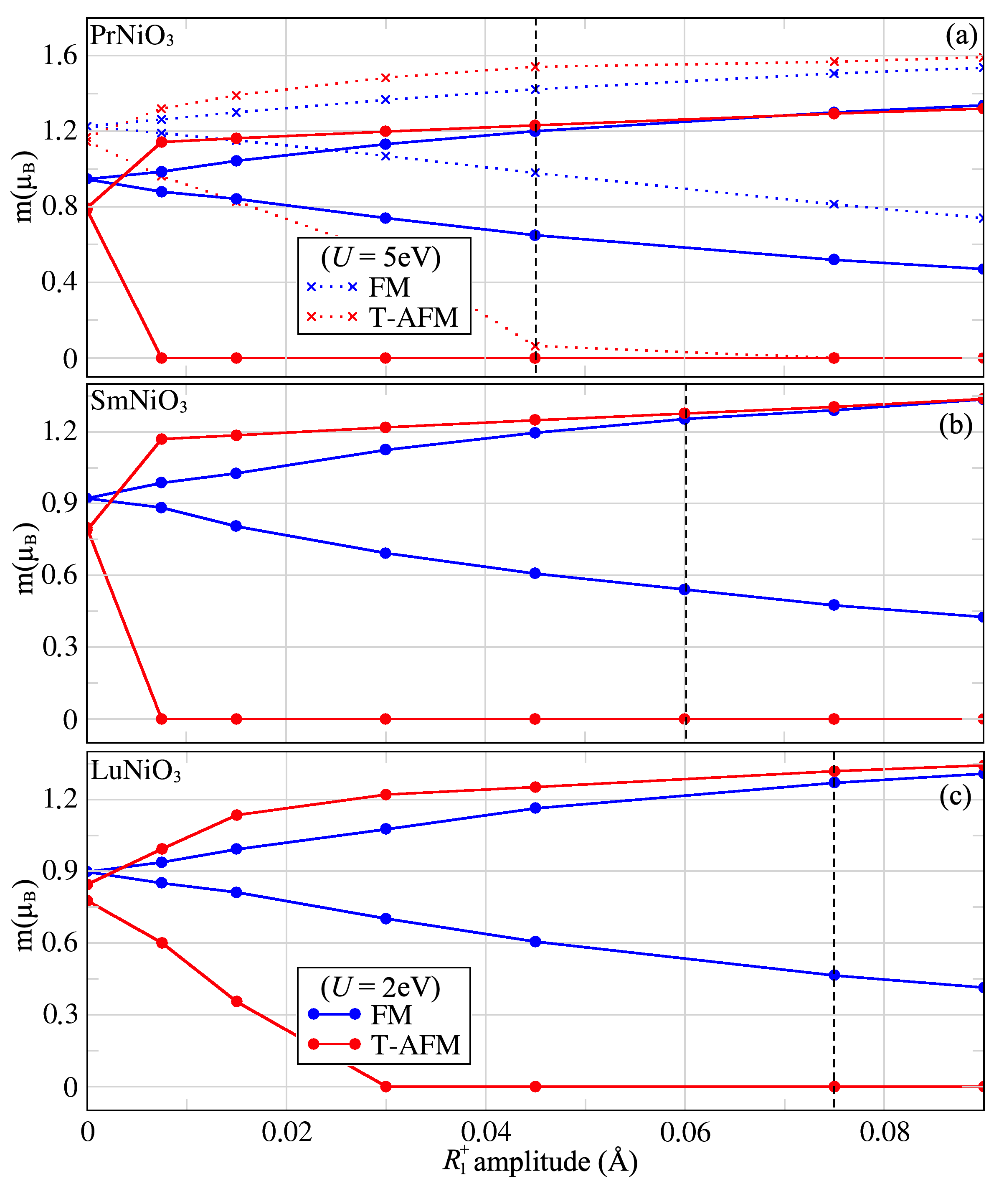}
    \caption{Evolution of the SB and LB magnetic moments of (a) PrNiO$_3$, (b) SmNiO$_3$, and (c) LuNiO$_3$ as a function of the \rone{} amplitude, calculated for different magnetic states using  $U$ = 2 eV.  Dashed vertical lines correspond to experimentally observed breathing mode amplitudes  from Ref.~\citenum{medarde2008} for PrNiO$_3$ and Ref.~\citenum{alonso2001} for LuNiO$_3$, and the theoretically predicted amplitude for SmNiO$_3$ from Ref.~\citenum{hampel2019}.   The \rone{} = 0 structure corresponds to the high temperature orthorhombic $Pbnm$ structure. Additional dotted curves for PrNiO$_3$ denoted by crosses represent results obtained with $U$ = 5 eV.
    }
    \label{fig:magnetic_moments}
\end{figure}

Nickel magnetic moments versus \rone{} amplitude are shown in Fig.~\ref{fig:magnetic_moments}(a)-(c). Dashed vertical lines denote the experimental breathing-mode amplitudes for PrNiO$_3$ (0.045 \AA \cite{medarde2008}) and  LuNiO$_3$ (0.075 \AA \cite{alonso2001}), and the theoretically predicted amplitude for SmNiO$_3$ (0.060 \AA \cite{hampel2019}), where no experimental data on the low temperature phase is available. In the case of the high-temperature orthorhombic phase (\rone{} = 0 \AA{}), in which all Ni octahedra have the same volume, FM order leads to magnetic moments of $\simeq 0.9 \, \mu_B$ per nickel atom  for all materials.  Within the T-AFM order the $Pbnm$ symmetry is broken, causing a small variation of magnetic moments for the inequivalent Ni sites. The difference between the SB and LB moments strongly increases with larger $R_1^+$ breathing mode amplitudes. In the FM case,  the SB Ni  magnetic moment smoothly decreases; for T-AFM, the moment quickly vanishes  at a modest value of \rone{}. The amplitude where the SB moment goes to zero depends on the rare-earth element, specifically it is larger for LuNiO$_3$ than PrNiO$_3$ and SmNiO$_3$. In all cases, the $R_1^+$ amplitudes at which the SB moment vanish are much smaller than the experimentally observed amplitudes for these compounds. Therefore, DFT+$U$ predicts $m_{\rm SB} = 0$ for T-AFM order for all experimental structures.  In contrast, $m_\text{LB}$ increases with \rone{} and converges to $\simeq 1.3  \, \mu_B$ for both FM and T-AFM order. 

The Wannier-based analysis allows us to analyze the orbital contributions to total nickel magnetic moments, and we find that for both FM and T-AFM order, LB and SB moments  originate almost entirely from Ni $e_g$ states. Since the corresponding Ni $t_{2g}$ states are fully occupied, they are not magnetically active.

These obtained results are consistent with the picture that would emerge from an analysis of the magnetization within the PAW spheres around Ni and O. The nickel moment from the Wannier analysis (and its dependence on \rone{}) is in good agreement ($\sim0.01$ $\mu_B$) with the moments in Ni PAW spheres. This indicates that the magnetism is localized around the atoms, instead of in the interstitial regions~\cite{danis_skyrmions}. Similarly to the PAW estimation, the Wannier based oxygen magnetic moments for all magnetic configurations are very small (less than 0.05 $\mu_B$). This suggests that, while there is significant Ni-O hybridization, the moment resides only at the Ni sites.

\begin{figure}[!t]
    \centering
     \includegraphics[width=1\linewidth]{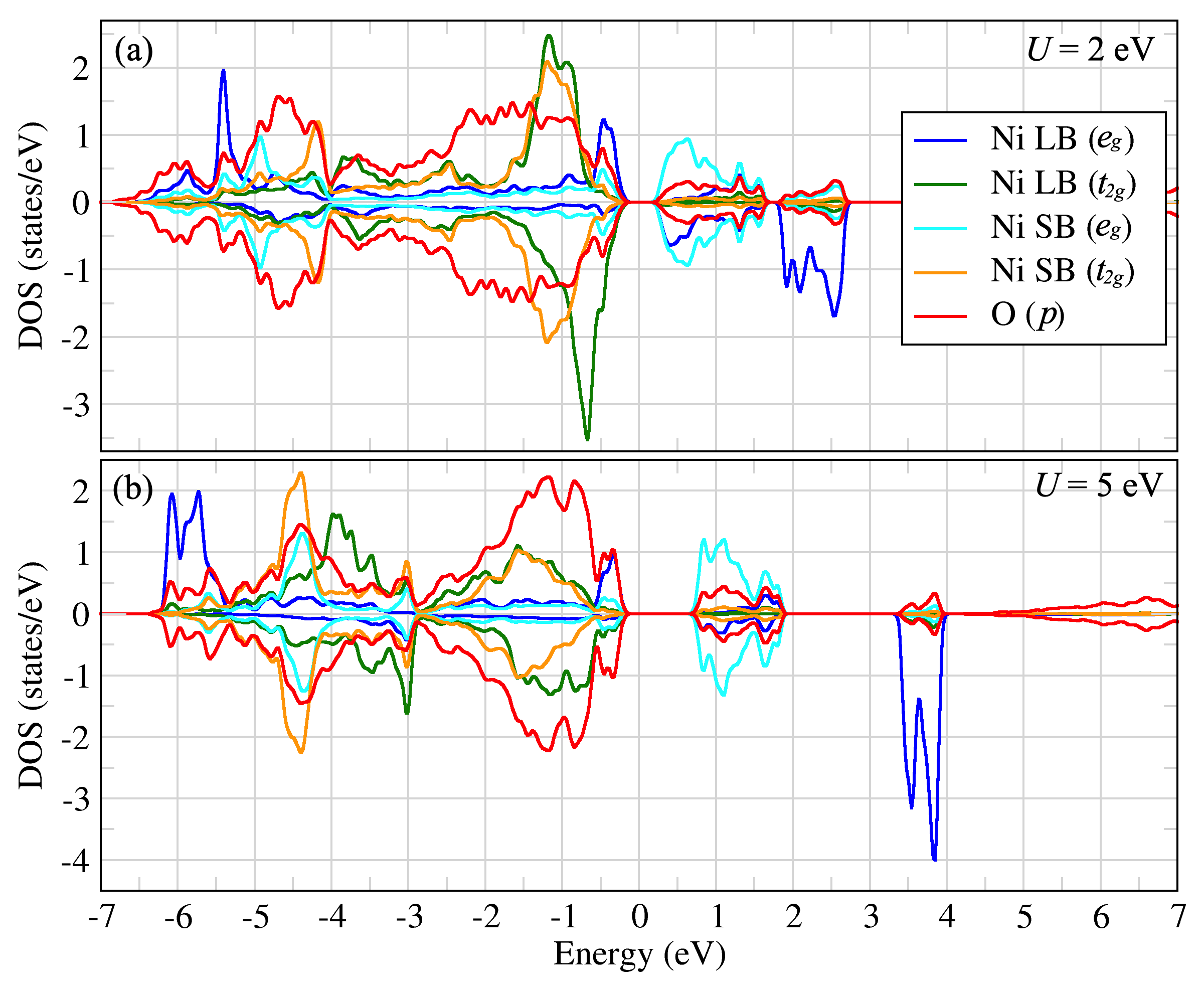}
    \caption{Partial densities of states (DOS) for PrNiO$_3$ with the experimental \rone{} = 0.045 \AA, calculated with T-AFM order for (a) $U$ = 2 eV and (b) $U$ = 5 eV. Here, only $t_{2g}$ and $e_{g}$ states contributions of a single LB and SB nickel atom, as well as  2$p$ states of the surrounding six oxygen atoms in NiO$_6$ octahedra are shown. The different occupation of the $e_g$ spin channels of the LB nickel site carries the magnetic moment, while for SB both spin channels are equally occupied, leading to a zero moment.  
    }
    \label{fig:DOS}
\end{figure}

The magnitude of the magnetic moments depend on the choice of  $U$ parameter of  DFT+$U$. Increasing $U$ leads to a stronger  localization of the $3d$ electrons, and consequently gives comparably larger total magnetic moments of nickel atoms. The reason for this can be seen in the density of states, e.g., for PrNiO$_3$  (Fig.~\ref{fig:DOS}), where a larger $U$ value leads to stronger splitting between spin channels of LB nickel $e_g$ states, enhancing their magnetic  moment. To demonstrate this  effect, we show magnetic moments calculated using  $U$ = 5 eV  for PrNiO$_3$ [see crosses in Fig.~\ref{fig:magnetic_moments}(a)]. The resulting magnetic moments for $Pbnm$ are larger than those obtained with  $U$ = 2 eV. For instance, the large \rone{} amplitude LB moments (for both magnetic orders) converge to $\simeq 1.3  \, \mu_B$ for $U$ = 2 eV versus $\simeq 1.5  \, \mu_B$ for $U$ = 5 eV. Furthermore,  $m_{\text{SB}}$ for T-AFM order becomes zero at comparably larger value \rone{} $\simeq$ 0.05 \AA \, compared to $\leq$ 0.007 \AA \, for $U$ = 2 eV, as was found in previous DFT+$U$ studies~\cite{hampel2017}.

\subsection{Exchange couplings}

\begin{figure}[!t]
    \centering
     \includegraphics[width=1\linewidth]{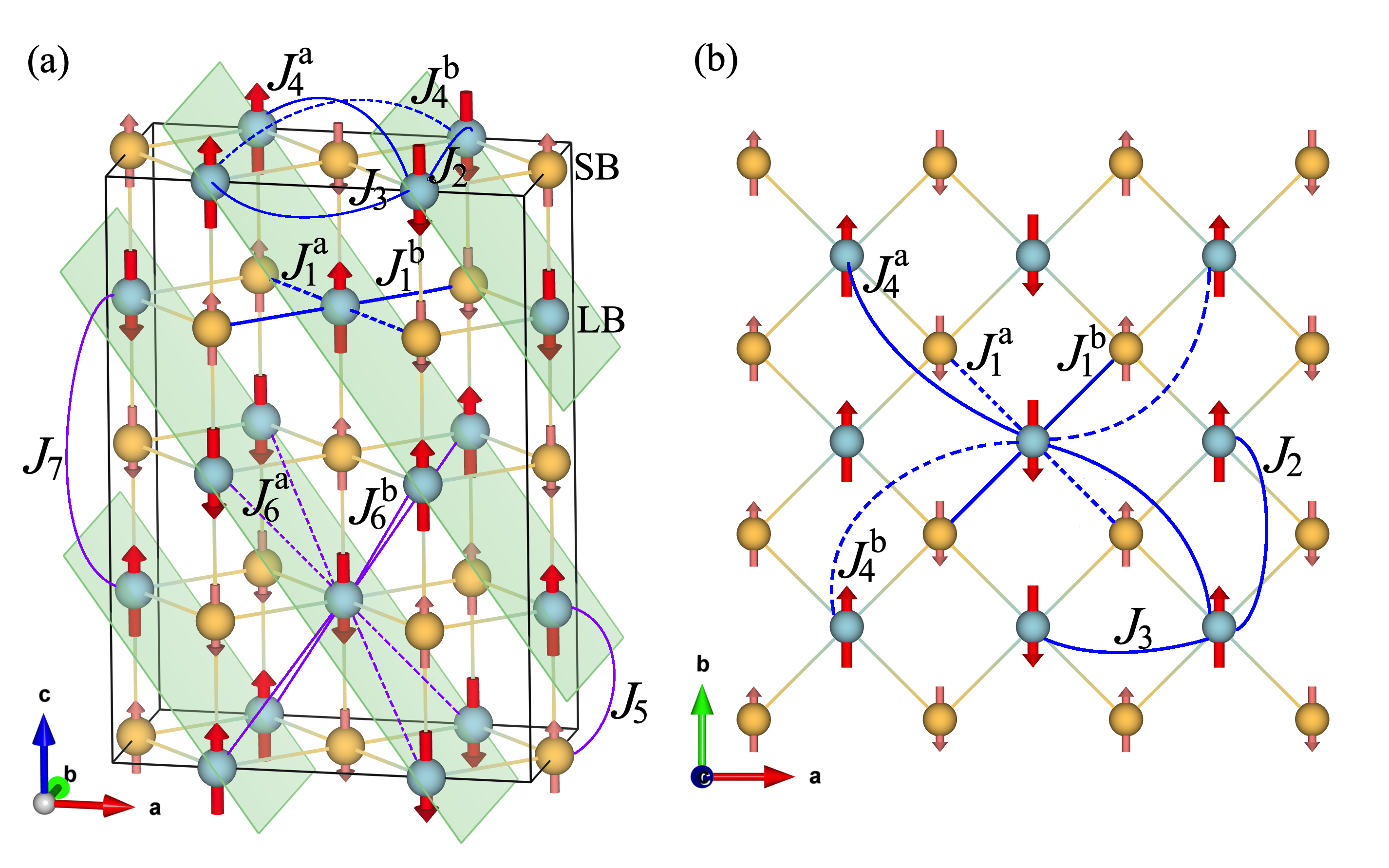}
    \caption{ Magnetic model of $\mathcal{R}$NiO$_3$ system with main exchange interactions plotted in (a) side and (b) top view. Blue atoms are long bond (LB) Ni, yellow are short bond (SB) (all other atoms removed). Arrows represent magnetic moments of nickel atoms  in the T-AFM order. Corresponding interactions between SB nickels are also included, but not shown in the figure for clarity. 
    }
    \label{fig:Model}
\end{figure}

We now perform a quantitative analysis of the various magnetic interactions present in the $\mathcal{R}$NiO$_3$ materials with respect to \rone{} amplitude. In Fig.~\ref{fig:Model} we show a schematic of the exchange interactions included in our study. The LB and SB Ni sublattices are depicted as blue and yellow spheres, and the RE and oxygens are removed for clarity. Focusing on a LB site, there are two inequivalent nearest-neighbor interactions with SB sites in the $ab$ plane ($J_1$). Also in the $ab$ plane, we consider next-nearest-neighbor interactions with other LB sites $J_2$ and $J_3$, as well as next-next-nearest neighbor interactions $J_4$. In the $c$ direction, we  have nearest neighbor interactions between LB and SB sites ($J_5$), next-nearest neighbor ($J_7$ and $J_6$) interactions.

\subsubsection{Ferromagnetic interactions $J_1$ and $J_5$ \label{sec:J1J5}}

\begin{figure}[!t]
    \centering
     \includegraphics[width=1\linewidth]{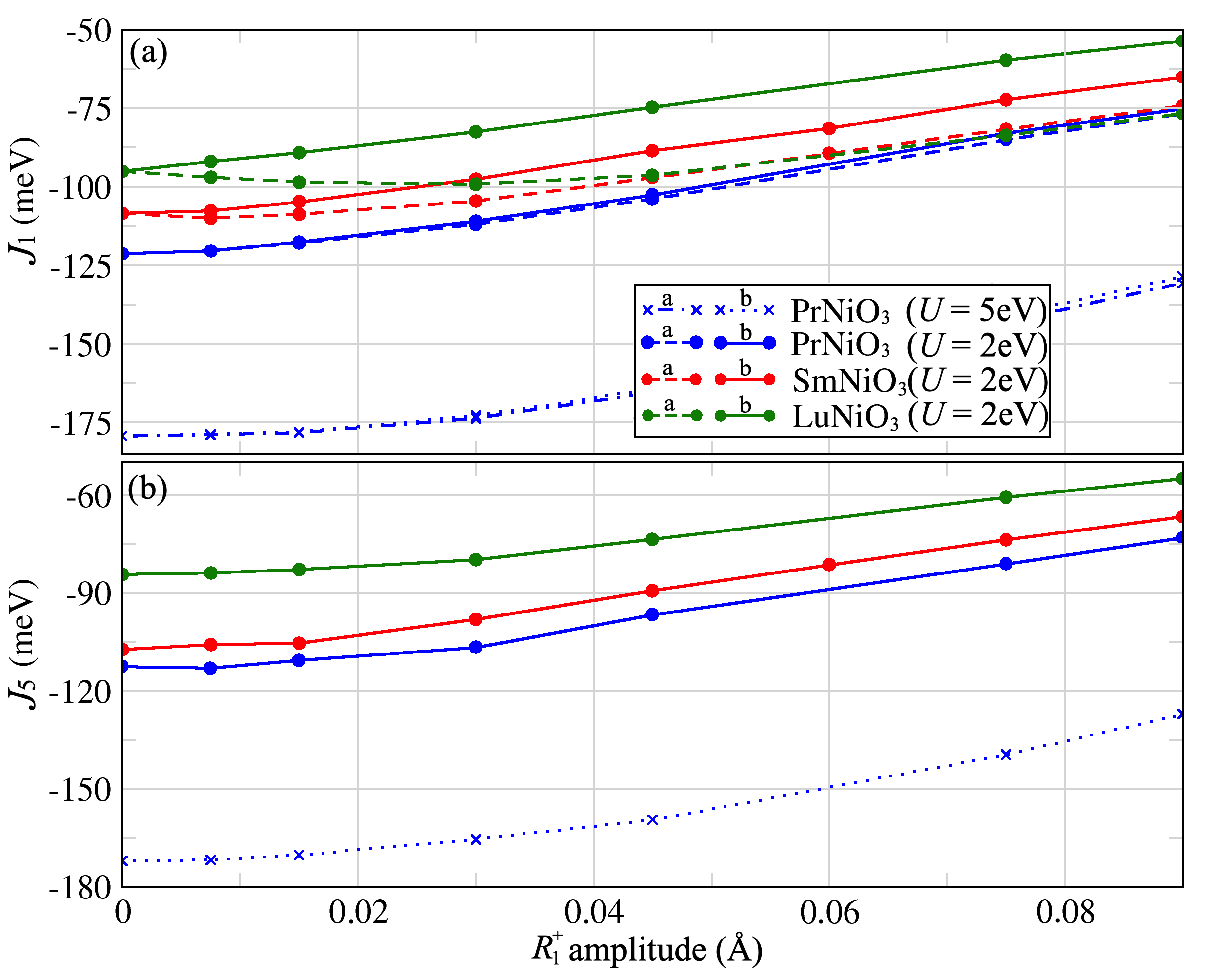}
    \caption{Nearest neighbor (a) $J_1^{a/b}$ and (b) $J_5$ exchange couplings (see Fig.~\ref{fig:Model}) between nickel atoms in $\mathcal{R}$NiO$_3$ as a function of breathing mode distortion \rone{} amplitude. The reference magnetic configuration is FM. The additional dotted  curves for PrNiO$_3$ denoted by crosses represent results obtained  with $U$ = 5 eV.  
    }
    \label{fig:J1-5}
\end{figure}

The calculated magnetic interactions via Eq.~(\ref{eq:Exchange}) depend on the reference magnetic configuration used in the underlying DFT calculation~\cite{zhu2020}, since the electronic structure and the resulting single-particle Green's functions will be different for FM and T-AFM order. Since the T-AFM order leads to a suppression of the SB magnetic moment (Fig.~\ref{fig:magnetic_moments}) the  nearest-neighbor exchange interactions  $J_1$ and $J_5$ are suppressed, as they couple two different types of nickel sites in the $ab$ plane and along the $c$ direction (Fig.~\ref{fig:Model}). Therefore,  for these interactions  we focus on the FM state as the reference magnetic order. Resulting values as a function of \rone{} are given in Fig.~\ref{fig:J1-5}.  

In general, short range interactions originate from a combination of different mechanisms~\cite{danis2016, danis2019}. Previous studies reported that the charge fluctuation between SB Ni and LB Ni sites with unequal spins will lead to ferromagnetic double exchange $J^{\text{DE}}<0$ process, which competes with the conventional antiferromagnetic superexchange $J^{\text{SE}}>0$~\cite{lu2018}. The resulting coupling $J = J^{\text{SE}} + J^{\text{DE}} $  can have  either sign depending on which mechanism will prevail. We show in Fig.~\ref{fig:J1-5} strong FM interaction (i.e., negative values) for both $J_1$ and $J_5$, implying  $\vert J^{\text{SE}} \vert < \vert J^{\text{DE}}\vert $ for both cases. Double exchange between the Ni atoms requires inequivalent charges on the sites~\cite{Zener_DE, Keshavarz_DE}. The strength of double exchange, and thus the FM couplings, would be expected to increasing with \rone{}~\cite{lu2018}. However, in our calculations  the magnitude of $J_1$ and $J_5$  couplings gradually \emph{decrease} with increasing \rone{} amplitudes. This suggests that other factors (e.g., charge transfer to oxygen states) are important in the resulting exchange interactions.

As a result of the presence of $M_3^{+}$ and  $R_4^{+}$ rotation modes, the nearest neighbor exchange $J_1$ splits into two groups $J^a_1$ and $J^b_1$ for finite \rone{} (Fig.~\ref{fig:Model}). The value of this splitting depends on the rare-earth element $\mathcal{R}$ (c.f., solid and dashed curves in Fig.~\ref{fig:J1-5}). In particular, a stronger splitting $J^a_1 < J^b_1$ can be found  for LuNiO$_3$  compared to PrNiO$_3$, where  $J^a_1 \simeq J^b_1$. Hence, this magnitude of the splitting is connected to the magnitude of the octahedral rotation modes $M_3^{+}$ and  $R_4^{+}$~\cite{hampel2017}.  As mentioned in Sec.~\ref{sec:comp}, RE elements with smaller effective radii will cause stronger $M_3^{+}$ and $R_4^{+}$ rotations of NiO$_6$ octahedra, modifying  Ni -- O -- Ni angles, and resulting in a larger difference between $J^a_1$ and $J^b_1$.

Furthermore, higher $U$ parameters will lead to a stronger localization of $d$ electrons, which influences not only the magnetic moments (Fig.~\ref{fig:magnetic_moments}), but also the corresponding exchange interactions between them. Our calculations  on PrNiO$_3$ with $U$ = 5 eV show an enhancement of the nearest neighbor ferromagnetic $J_1$ and $J_5$ couplings (Fig.~\ref{fig:J1-5}). This is directly related to the fact that  using a larger on-site Coulomb interaction parameter  will decrease the AFM superexchange term, which is inversely proportional to $U$.  Double exchange is proportional to the transfer integral between Ni site, and does not depend directly on $U$~\cite{Keshavarz_DE}. 

\subsubsection{ Antiferromagnetic interactions $J_4$ and $J_7$}

\begin{figure}[!t]
    \centering
     \includegraphics[width=1\linewidth]{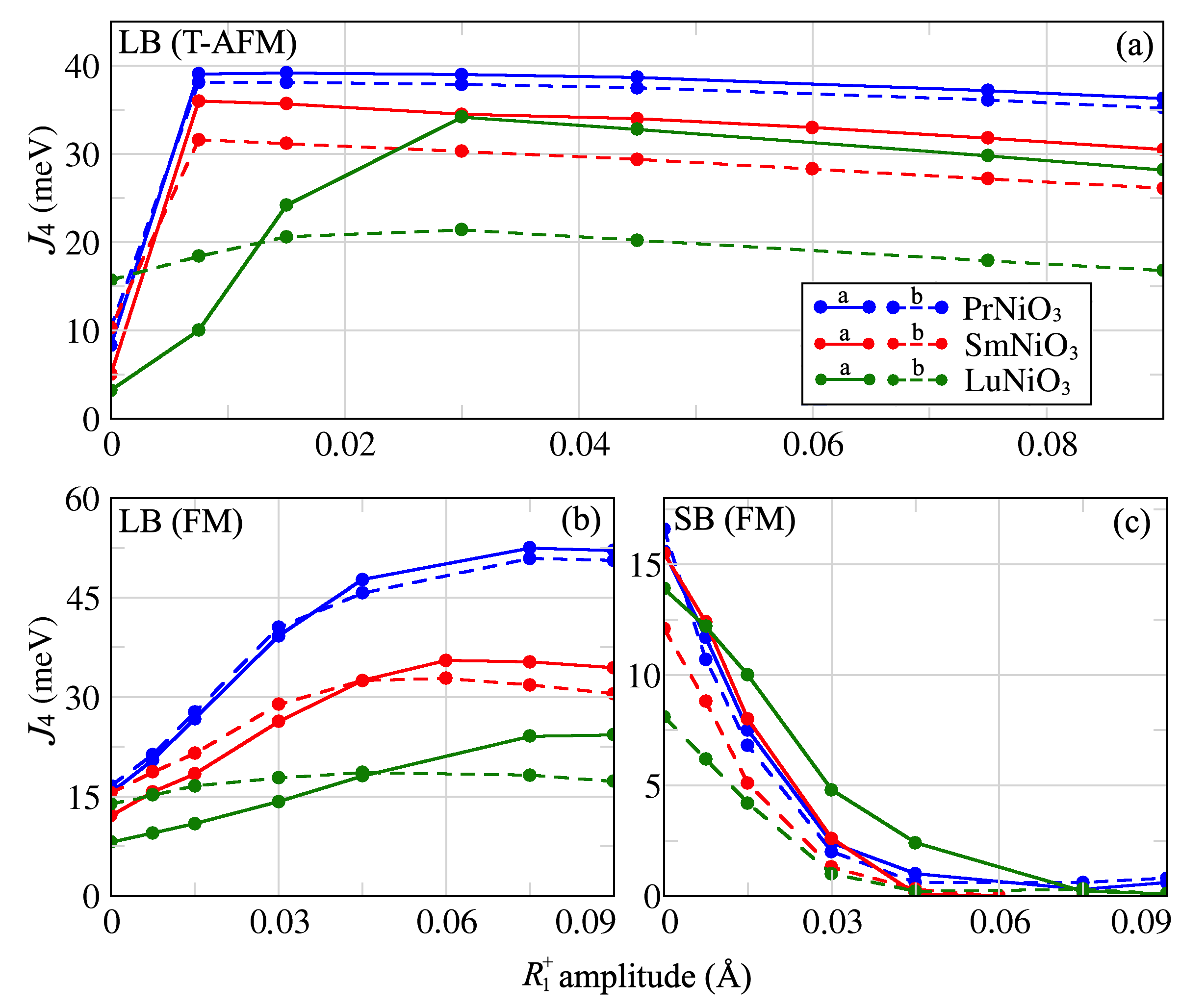}
    \caption{ Next nearest neighbor $J_4^{a/b}$ magnetic interactions between nickel atoms in $\mathcal{R}$NiO$_3$ as a function of breathing mode distortion \rone{} and reference magnetic order: (a) interactions between LB atoms in T-AFM order, while (b) and (c) correspond to interactions between LB and SB in FM order, respectively. Notation of magnetic couplings corresponds to  Fig.~\ref{fig:Model}. 
    }
    \label{fig:J4}
\end{figure}

\begin{figure}[!t]
    \centering
     \includegraphics[width=1\linewidth]{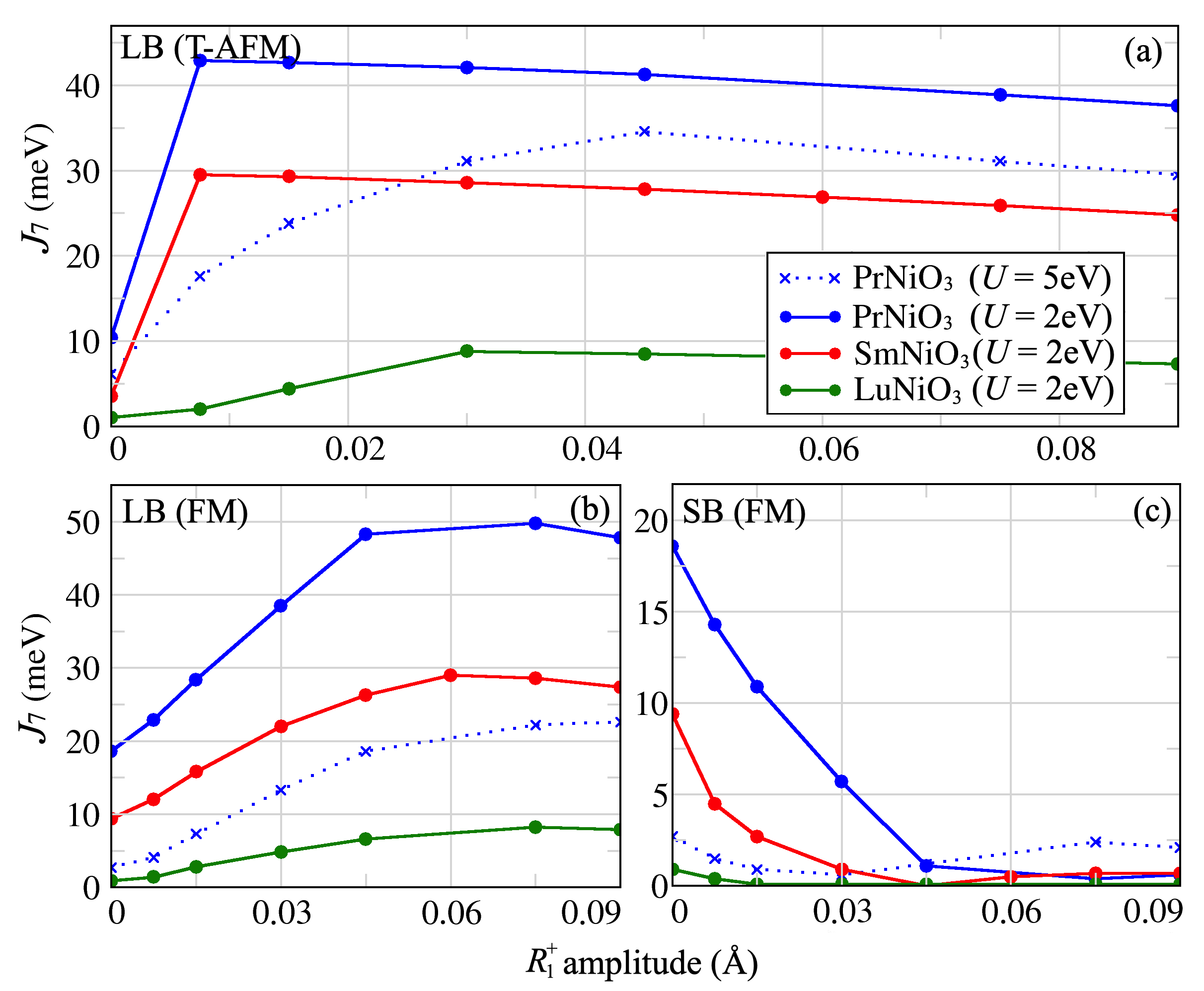}
    \caption{ Next nearest neighbor $J_7$ magnetic interactions between nickel atoms in $\mathcal{R}$NiO$_3$  as a function of breathing mode distortion \rone{} and reference magnetic order: (a) interactions between LB atoms in T-AFM order, while (b) and (c) correspond to interactions between LB and SB in FM order, respectively. Additional dotted  curve for  PrNiO$_3$ denoted by crosses  represents results obtained with $U$ = 5 eV. Notation of magnetic couplings corresponds to  Fig.~\ref{fig:Model}.  
    }
    \label{fig:J7}
\end{figure}

Next-nearest-neighbor exchange interactions $J_4$ and $J_7$ connect the same type (SB or LB) of nickel atom (Fig.~\ref{fig:Model}). Similar to the nearest-neighbor $J_1$ interaction, the next nearest neighbor $J_4$ also splits into two groups  $J^a_4$ and $J^b_4$. The value of this splitting depends on the RE element via the  octahedral rotation modes $M_3^{+}$ and  $R_4^{+}$. These interactions are antiferromagnetic, i.e., positive  (Figs.~\ref{fig:J4} and \ref{fig:J7}), and  the dependence on the breathing mode distortion is different depending on the reference magnetic order. 

For the case of FM order, the next-nearest neighbor exchange interactions between LB nickel atoms gradually increase, while for the SB nickels they gradually decrease to zero; this is the same trend as the magnetic moments, see Fig.~\ref{fig:magnetic_moments}. In contrast, for the T-AFM order, the SB magnetic moments are suppressed and thus the magnetic interactions between LB nickel atoms jump abruptly to a large value with $R_1^+>0$. Further increasing \rone{} results in a small reduction in the magnitude of the exchange coupling. The behavior in Fig.~\ref{fig:J4}(a) hints at a strong coupling between \rone{} distortion and T-AFM  order. Overall, exchange couplings between LB nickel sites at both reference magnetic configurations have the same order of magnitude. The values of $J_4$ and $J_7$ at $R_1^+$ = 0 are about an order of magnitude smaller than $J_1$/$J_5$, and thus the  T-AFM magnetic configuration is fragile. Since the main mechanism of the next nearest interactions is superexchange, all these couplings weaken for a larger $U$ value, which is shown for $J_7$ of PrNiO$_3$ in Fig.~\ref{fig:J7}.

\subsubsection{Summary and remaining exchange couplings}

\begin{table}[t]
\centering
\caption {The main exchange couplings (in meV) for PrNiO$_3$, SmNiO$_3$ and LuNiO$_3$, calculated ($U$=2 eV) using experimental breathing mode amplitudes \rone{} for PrNiO$_3$ and LuNiO$_3$, and the theoretical one for SmNiO$_3$.  Notation of magnetic couplings corresponds to  Fig.~\ref{fig:Model}. }
\begin{ruledtabular}
\begin {tabular}{c|ccc}
             & PrNiO$_3$ & SmNiO$_3$   &  LuNiO$_3$  \\\hline
   \rone{}          &  0.045 \AA~\cite{medarde2008} &  0.060 \AA~\cite{hampel2019}    & 0.075 \AA~\citep{alonso2001}  \\
              \hline 
 $J_{1a}/ J_{1b}$ &  -103.9/-102.6    & -89.4/-81.4 & -83.6/-59.8 \\
 $J_5$  & -96.7 & -81.4 & -60.7 \\
                \hline      
 $J_{4a}/ J_{4b}$ &     38.7/37.5 &   33.0/28.3   & 29.8/17.9 \\
 $J_7$ & 41.3 & 26.9 & 7.7 \\
                 \hline 
  $J_2$ & 3.5  & 7.5 & 10.0 \\
  $J_3$ & 9.9  & 11.6 & 16.0 \\
   $J_{6a}/ J_{6b}$ &    10.9/7.2 &       7.8/7.7   & 6.2/8.5 \\                            

\end {tabular}
\end{ruledtabular}
\label{tab:Exchange_couplings}
\end {table}

The values for the exchange interactions discussed above, calculated at the experimental $R_1^+$ amplitudes \cite{medarde2008,alonso2001} (except for SmNiO$_3$ where we use the theoretical value from Ref.~\onlinecite{hampel2019}), are given in Table~\ref{tab:Exchange_couplings}. Also included are the remaining couplings shown in Fig.~\ref{fig:Model}: $J_2$, $J_3$, and $J_6$; we see that these couplings are small compared to $J_1/ J_5$ and $J_4/ J_7$. This is because of the geometry of the active $e_g$ orbitals,  which overlap with each other via oxygen atoms along the Cartesian axes (Fig.~\ref{fig:Model}). Our orbital resolved analysis of exchange interactions (Eq.~\ref{eq:Exchange}) demonstrates that the main contribution to in-plane $J_1$ and $J_4$ comes from $d_{x^2- y^2}$ orbitals, while out-of-plane $J_5$ and $J_7$ are originated from $d_{z^2}$ states. Furthermore, it can be seen that the exchange couplings significantly increase going from Lu to Pr, e.g. when octahedral rotations are reduced.

\subsection{Anisotropy and non-collinear order \label{sec:ncl}}

To shed light on the long-standing question of collinear versus noncollinear (NCL) order in rare-earth nickelates (Fig.~\ref{fig:Magnetic_ordering}), we also performed an analysis of the magnetic anisotropy and NCL magnetic arrangements within DFT+$U$ calculations taking spin-orbit coupling (SOC)  into account.  In the case of the proposed \cite{scagnoli2006, lu2018} NCL magnetic order, nearest-neighbor magnetic moments are orthogonal to each other [Fig.~\ref{fig:Magnetic_ordering}(b)]. Since  $\mathbf{m}_{\rm LB} \cdot \mathbf{m}_{\rm SB}  = 0$, the contributions of the strong isotropic ferromagnetic  $J_1/ J_5$ couplings are eliminated, and magnetic frustration between FM nearest-neighbors and AFM next-nearest-neighbors couplings does not occur. 

We perform calculations including SOC on the $Pbnm$ crystal structure of PrNiO$_3$, SmNiO$_3$, and LuNiO$_3$ with  breathing mode amplitudes given in Table~\ref{tab:Exchange_couplings}.  We find that the SB moments for all three materials vanishes in these calculations, analogously to the collinear calculations in  Fig.~\ref{fig:magnetic_moments}. The LB moment is always nonzero and equals 1.19 $\mu_B$, 1.24 $\mu_B$ and 1.28 $\mu_B$ for $\mathcal{R}$ =  Pr, Sm and Lu, respectively. Thus, the  NCL configuration we find is equivalent to the collinear T-AFM order, with the LB magnetic moments free to align  along the direction which minimizes the anisotropy energy in the systems; we can find this direction by calculating the anisotropic energy map~\cite{abdeldaim2019}.    

\begin{figure}[!t]
    \centering
     \includegraphics[width=1\linewidth]{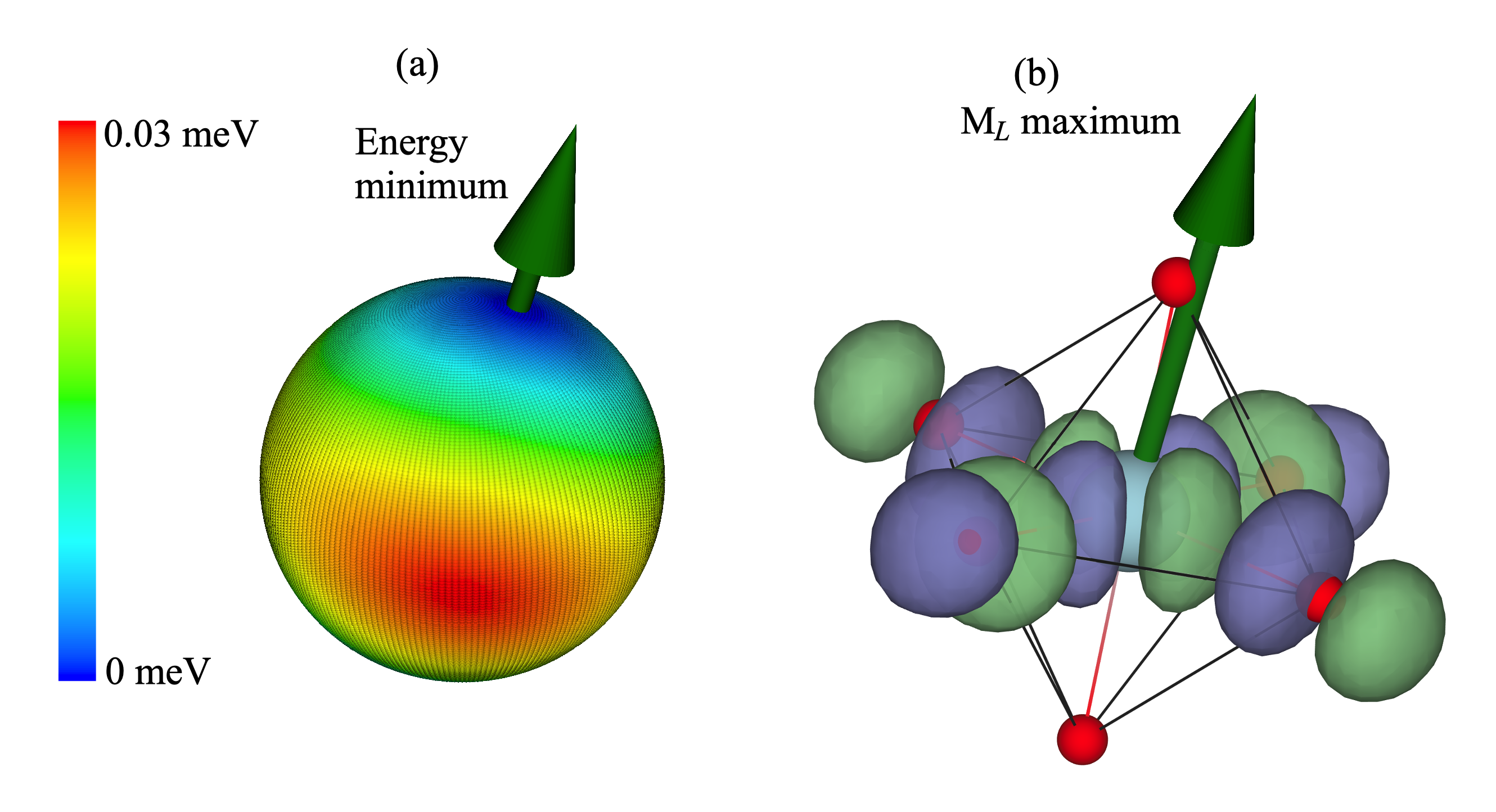}
    \caption{Anisotropic energy map for PrNiO$_3$ system. The arrow indicates the energy minimum, which is close to the direction of apical oxygen position of MoO$_6$ octahedra, depicted in (b) together with Wannier function of $x^2 -y^2$ symmetry, the valence electron in which state induces orbital magnetic moment. 
    }
    \label{fig:Anisotropy}
\end{figure}

In Fig.~\ref{fig:Anisotropy}(a), we show the interpolated anisotropic energy map for PrNiO$_3$. To generate this plot, DFT+$U$+SOC calculations were performed using magnetic moments constrained along various (uniformly distributed) directions, but keeping their mutual orientation consistent with T-AFM order. We find an easy axis  along a direction close to the position of the apical oxygen atom in the NiO$_6$ octahedra [green arrow in Fig.~\ref{fig:Anisotropy}(a)]. The anisotropy in the energy is very small ($\sim$ 0.03 meV), which is expected for $3d$ systems with weak SOC~\cite{danis2016, danis2019}.  According to conventional theories of magnetocrystalline anisotropy (i.e., Bruno’s model~\cite{bruno1989, Bruno_rule}), the orbital moment is maximized when the magnetization points along the easy axis. This is true for $\mathcal{R}$NiO$_3$ systems, showing the maximum value of orbital moment $M_{L}$ = 0.13 $\mu_B$  aligned with spin magnetic moment 1.19 $\mu_B$ along the direction of the easy axis.

The orbital moment is induced by $e_g$ electrons, which can be seen on the level of density matrix $\hat{N}$:
\begin{eqnarray*}
M_L^{z} = {\rm Tr}_{m} (\hat{N} \cdot \hat{L}_{z}).  
\label{eq:Orbital moment}
\end{eqnarray*}
In particular, the  valence electron occupying the $d_{z^2}$ will not produce any orbital moment ($\hat{L}_{z}d_{z^2} = 0$). On the other hand, due to a non-zero $\braket{d_{xy}|\hat{L}_z|d_{x^2-y^2}} = 2i$, and the fact that the calculated density matrix contains imaginary components between $d_{xy}$ and $d_{x^2-y^2}$ states, an orbital moment is induced along the direction of the apical oxygen in the NiO$_6$ octahedra ($z$-axis) by the electron in the $d_{x^2-y^2}$ state. This leads to the formation of the easy axis anisotropy found by in our DFT+$U$+SOC calculations (Fig.~\ref{fig:Anisotropy}).

\section{Discussion}
\label{sec:discussion}

The goal of the calculations in Sec.~\ref{sec:results} was to elucidate two key issues regarding the magnetic structure in RE nickelates: (a) why DFT+$U$ predicts a FM structure, while an AFM structure with propagation vector $\mathbf{q} = (\frac{1}{4}, \frac{1}{4},\frac{1}{4})$ is observed experimentally; and (b) what is the relation between the T-AFM magnetic structure determined from neutron diffraction data, and the NCL magnetic structure determined from RIXS.

We find that the key physical mechanism at play in these systems is the competition between strong nearest-neighbor ferromagnetic $J_1/ J_5$ and the much weaker next-nearest-neighbor antiferromagnetic $J_4/ J_7$. This competition causes magnetic frustration, and works to destabilize the long-range magnetic order. As discussed before, the \rone{} distortion will result in different magnetic moments on the LB and SB Ni atoms, with the SB magnetic moment suppressed (Fig.~\ref{fig:magnetic_moments}). Therefore, the energy gain from forming ferromagnetic order will be directly proportional to the SB moment. Hence, the reason that DFT+$U$ favors FM order~\cite{hampel2017,varignon2017} is the overestimation of short bond magnetic moments, leading to a larger contribution of nearest-neighbor FM couplings, which makes the FM solution energetically favorable. For a sufficiently small SB moment the FM $J_1$ and $J_5$ interactions are suppressed, and the next-nearest neighbor AFM couplings $J_4$ and $J_7$ will dominate; the result is the collinear T-AFM order proposed from neutron diffraction experiments~\cite{garcia1994, alonso1999, Munoz:2009, gawryluk2019} (Fig.~\ref{fig:Magnetic_ordering}). As we saw in Fig.~\ref{fig:magnetic_moments}, the SB moments go to zero rapidly with increasing \rone{} amplitude in the T-AFM ordering to avoid the energy penalty associated with a finite FM  $J_1/J_5$ coupling. 

The reason for the theory-experiment discrepancy, could be due to an overestimation of exchange splitting in DFT+$U$, or problems in the refinement of experimental data. For a future analysis it would be of high interest to incorporate self-energy corrections to the Green's function exchange coupling analysis from higher order theory, i.e, GW or dynamical mean-field theory, to rule out problems due to dynamic correlation effects~\cite{Kvashnin2015}.

Another approach to avoiding frustration is the proposed magnetic structure from RIXS experiments \cite{scagnoli2006, lu2018}. The NCL ordering where the nearest-neighbor moments point in orthogonal directions, has no  energy penalty for AFM versus FM ordering, and thus does not require the strong reduction of the SB magnetic moment needed for stabilization of collinear antiferromagnetic order. However, as discussed in Sec.~\ref{sec:ncl}, we cannot stabilize this ordering in our calculations, as the SB moment vanishes at the experimental \rone{} amplitude, even if a SOC is included (thus a NCL order is allowed). The origin of the SB magnetic suppression can be related to the symmetry of the magnetic configurations. The SB magnetic moment is surrounded by four LB nickel atoms within the $ab$ plane (Fig.~\ref{fig:Magnetic_ordering}), and in the case of FM order, this will enhance local magnetic polarization on the SB nickel atom, increasing its magnetic moment. But since in T-AFM (and also the NCL order) half of the neighboring LB moments are aligned parallel, and the other half anti-parallel, the SB moment will be reduced. The valence electron in the magnetically active $d_{x^2-y^2}$ states induces an orbital moment $M_L \simeq$ 0.13 $\mu_B$ pointed at the apical oxygen position of the NiO$_6$ octahedra. This leads to the formation of easy-axis anisotropy, giving preferable direction for LB magnetic moments alignment.

The analysis above is for the experimental \rone{} amplitudes (theoretical for SmNiO$_3$); however, we see that these results are robust over a wide range of \rone{} amplitudes. Increasing the breathing distortion reduces FM nearest neighbor $J_1$/ $J_5$ interactions and magnetic moments of SB nickel atoms, thus it can strengthen long range T-AFM ordering in these systems. The same behavior was found in previous calculations~\cite{hampel2017} by directly comparing relaxed \rone{} amplitudes in DFT+$U$, which revealed \rone{} (FM) $ < $ \rone{}(T-AFM). 

The experimental estimates, used to fit magnon branches in RIXS experiment for thin film NdNiO$_3$ samples, revealed that the in $ab$-plane next-nearest-neighbor interaction $J_4$ is stronger that the nearest neighbor $J_2$ coupling $J_4 \simeq 2J_2$~\cite{lu2018}. In our calculations we also observe  $J_4 \gg J_2$ (Table~\ref{tab:Exchange_couplings}). Furthermore, we found strong spacial anisotropy depending on RE element, which splits interactions in the $ab$ plane into two groups: nearest-neighbor $J_{1a}/J_{1b}$, next-nearest-neighbor $J_{2}/J_{3}$ and next-next-nearest neighbor $J_{4a}/J_{4b}$ (Fig.~\ref{fig:Model}). Moreover, we see a strong difference between $ab$ plane ($J_1/J_{4}$) and out-of-plane $c$-direction ($J_5/J_{7}$) exchange couplings depending on RE element. 

Finally, the results provide a clear picture of the trends of the magnetic order across the nickelate series. While the larger octhadral distortions ($R_4^+,M_3^+$) present for smaller RE ions make the system more susceptible to the insulating breathing-mode-distorted state~\cite{peil2019,mercy2017}, we find that they also reduce the exchange couplings, making the formation of the magnetically ordered state less favorable. This is in agreement with the experimentally-observed reduction in stability of magnetic order in the low temperature insulating phase across the series towards LuNiO$_3$~\cite{hampel2017, catalano2018, gawryluk2019}.

\section{Conclusions \label{sec:conc}}

Based on DFT+$U$ calculations, Wannier orbitals and Green's function techniques, we constructed microscopic magnetic models for selected rare-earth nickelate systems (PrNiO$_3$, SmNiO$_3$, and LuNiO$_3$) and studied the evolution of their parameters under the \rone{} breathing mode distortion and reference magnetic configuration. Our theoretical results help to clarify several issues related to these systems. The Wannier function based analysis demonstrates that the magnetic moments reside on nickel atoms, and they are induced by active Ni $e_g$ states. The main exchange interactions are strong ferromagnetic interactions arising from double exchange between the nearest neighbor nickels, and somewhat weaker antiferromagnetic interactions between the next-nearest neighbor nickels  induced by a kinetic superexchange mechanism. The competition between these, together with the \rone{} distortion, determines the magnetic order. 

The stronger nearest-neighbor interactions are the reason that DFT+$U$ calculations find the ferromagnetic state to be lowest in energy. The stabilization of collinear T-AFM order, which is compatible with the experimental symmetry, is possible due to suppression of short bond magnetic moments with the onset of the breathing mode distortion, eliminating the contribution of the ferromagnetic interactions. We find that even if  spin-orbit coupling is included and a non-collinear ordering is allowed, the SB Ni moments are still suppressed, and T-AFM order is restored with an easy axis magnetization pointed towards the apical oxygen. Thus our work determines the reason for the general disagreement between the DFT-$U$ predicted magnetic order (FM) and experiment (AFM), as well as provides a clear explanation how the $\mathbf{q} = (\frac{1}{4}, \frac{1}{4},\frac{1}{4})$ magnetic order emerges in rare-earth nickelates, and how the trends evolve across the series.

\acknowledgments
 D.I.B. is grateful to S. A. Nikolaev (Tokyo Institute of Technology)  for helpful discussions.  The calculations have been performed using the facilities of the Flatiron Institute. The Flatiron Institute is a division of the Simons Foundation. C.E.D. acknowledge support from the National Science Foundation under Grant No. DMR-1918455.



%

\end{document}